\documentstyle[floats,aps]{revtex}

\begin{document}
\draft
\catcode`\@=11
\catcode`\@=12
\twocolumn[\hsize\textwidth\columnwidth\hsize\csname%
@twocolumnfalse\endcsname
\title{The Microscopic Model of Composite Fermion Type Excitations 
for $\nu=\frac{1}{m}$ Edge States}
\author{Yue Yu and Zhongyuan Zhu}
\address{Institute of Theoretical Physics, Academia Sinica, Beijing 100080,
China}

\maketitle
\begin{abstract}
 
We derive a microscopic theory of the composite fermion type quasiparticles 
describing the low-lying edge excitations in the fractional quantum 
Hall liquid with $\nu=1/m$. Using the composite fermion
transformation, one finds that the edge states of the 
system in a disc sample are described by the
Calogero-Sutherland-like model (CSLM) in the one-dimensional limit. 
This result presents
the consistency between one-dimensional and two-dimensional statistics. 
It is shows that the low-lying excitations, indeed, have the chiral 
Luttinger liquid behaviors because there is a gap between 
the right- and left-moving excitations of the CSLM.    

\end{abstract}

\pacs{PACS numbers: 73.40.Hm,71.10.+x,71.27.+a}]

It is well-known that the bulk states of integer and fractional quantum Hall 
effects  (IQHE and FQHE), in some sense, are very like to insulator states
because there is an energy gap between the excitation state and the ground 
state. However, the transport behaviors of the quantum Hall states are 
remarkably different from those of the insulator. This difference leads to 
the recognization of the specific behavior of the quantum Hall gapless edge 
states \cite{Halp}. A  chiral Fermi liquid description to the edge states of 
IQHE has been provided by Halperin \cite{Halp}. However,  the Fermi liquid 
notion could not be applied to FQHE for either the bulk or the edge states 
since the strong correlations among the electrons in a high magnetic field. 
A very thoroughly investigation for the edge states of FQHE has been done by 
Wen from the chiral Luttinger liquid point of view \cite{Wen}. Recently, the
edge excitations of FQHE were studied by several groups numerically 
\cite{Rez} or in accordance with the Calogero-Sutherland-like model (CSLM) 
\cite{ECS} as well as the composite fermion (CF) picture in the Hartree 
approximation \cite{CFP}. In this Letter, however, we will give a microscopic
model of the CF-type excitations for the edge excitations while the bulk states
lie on the mean-field states. 
Starting from the Hamiltonian of the
two-dimensional interacting electrons in high magnetic field and using the
composite particle transformation \cite{KHZ,LF,KZ,HLR}, 
one transforms the system to a CF (or composite boson) system if the 
statistics parameter is chosen as an even integer $\tilde\phi$ (or an odd $m$)
. We would like to consider the low-lying excitations. 
For $\nu=1/m$, the bulk states are gapful quantum Hall liquids. So, they 
can be decoupled to the low-lying excitation sector of the quantum state space.
The only low-lying excitations are edge excitations. 
The edge of the system is considered as a circular strip with an infinitesimal
width near the boundary of the quantum Hall liquid droplet.  
We will begin with the CF model. A Fermi-liquid-like theory could be applied  
to the FQHE both of the bulk and edge states \cite{HLR,CFP}. Using the
Fermi liquid notion to the CF system, we have $N^e$ CF-type 
quasiparticles. Grafting Halperin's single particle
argument for the edge states of IQHE to the present case, one can get a
microscopic model of the CF-type excitations.
In the one dimension limit, we see that the edge excitations could be 
described by the CSLM \cite{CS} perturbed by the interaction between 
the CFs. One has argued before by one of us, cooperating 
with Wu, the CSLM can be thought as the fixed point of the Luttinger liquid as 
the ideal Fermi gas versus the Fermi liquid \cite{Wu}. This implies that the
edge state of the FQHE could be described by the Luttinger liquid. 
However, one shows that the right- and left- moving sectors opens a gap 
$|n|\hbar\omega^*_c$ because of the existence of the magnetic field. 
This implies that the edge excitations are chiral, 
which is consistent with the chiral
Luttinger liquid description of the edge excitations \cite{Wen}. 
Our result also gives the self-consistency between one-
and two-dimensional fractional statistics, which improves  
the observation that the CSLM could be mapped into an anyon model on a ring 
\cite{Li}. The short range interaction between the CFs will not effect the 
exponents.
The Coulomb interaction provides a branch of
charge plasma excitations.

The two-dimensional interacting electrons which are polarized by a high
magnetic field are governed by the following Hamiltonian
\begin{eqnarray}
H_{\rm el}&=&\sum_{\alpha=1}^N\frac{1}{2m_b}[\vec{p}_\alpha
-\frac{e}{c}\vec{A}(\vec{r}
_\alpha)]^2+\sum_{\alpha<\beta}V(\vec{r}_\alpha-\vec{r}_\beta)\\ \nonumber
&+&\sum_\alpha U(\vec{r}_\alpha),
\label{HO}
\end{eqnarray}
where
$V(\vec{r})$ is the interaction between electrons. 
$m_b$ is the band mass of the electron
and $U(\vec{r})$ is the external potential. The composite particle 
transformation will bring us to a good
starting point to involve in the FQHE physics as many successful investigations
told us \cite{KHZ,LF,KZ,HLR}. We begin with the CF 
transformation which reads
\begin{equation}
\Phi(z_1,...,z_N) = \prod_{\alpha<\beta}\biggl[\frac{z_\alpha-z_\beta}
{|z_\alpha-z_\beta|}\biggr]^{\tilde\phi} 
\Psi(z_1,...,z_N),
\label{CBT}
\end{equation}
where $\Phi$ is the electron wave function. The CF consists of an
electron attached by $\tilde\phi$ flux quanta.  
By using the CF theory, the bulk behavior of the FQHE has been 
well-understood \cite{Jain,HLR}.
We, now, would like to study the microscopic theory of the CF
edge excitations. The partition function of the system is given by
\begin{eqnarray}
Z&=&\sum_{N^e} C^{N^e}_{N}\int_{\partial} d^2z_1....d^2z_{N^e}
\int_{B}d^2z_{N^e+1}...d^2z_N\\ \nonumber
&\times&\biggl(\sum_\delta |\Psi_\delta|^2 e^{-\beta 
(E_\delta+E_g)}+\sum_\gamma|\Psi_\gamma
|^2e^{-\beta (E_\gamma+E_g)}\biggr),
\label{pf}
\end{eqnarray} 
where we have divided the sample into the edge $\partial$ and the bulk $B$. 
$E_g$ is the ground state energy and $E_\delta$ are the low-lying gapless 
excitation energies with $\delta$ being the excitation  branch index.
$E_\gamma$ are the gapful excitation energies. At $\nu=1/\tilde\phi$, the 
low-lying excitations are everywhere in the sample and we do not consider this
case here. We are interested in the case $\nu=\frac{1}{\tilde\phi+1}=1/m$,
where the bulk states are gapful. The low-lying excitations are confined in 
the edge of the sample. For convenience, we consider a disc geometry sample
here. The edge potential is postulated with a sharp shape. The advantage
of the CF picture is we have a manifestation that the FQHE
of the electrons in the external field $B$ could be understood as the IQHE of 
the CFs in the effective field $B^*$ defined by $B^*\nu^*=B\nu$.
For the present case, $B^*=B/m$ and $\nu^*=1$. The energy gap in the bulk
is of the order $\hbar \omega_c^*$ with the effective cyclotron frequency
$\omega_c^*=\frac{eB^*}{m^* c}$ ($m^*$ is the effective mass of the CF). 
Hereafter, we use the unit $\hbar=e/c=2m^*=1$ except the explicit expressions.
By the construction 
of the CF, the FQHE of the electrons can be described by  
the IQHE of the CFs \cite{Jain} while the electrons in the $\nu=
1/\tilde\phi$ field could be thought as the CFs in a zero effective field.
Thus, a Fermi-liquid like theory could be used \cite{HLR} and we 
have a set of CF-type quasiparticles. Applying the single
particle picture, which Halperin used to analyze the edge excitations of
the IQHE of the electrons, to the edge excitations of
the CFs, one could have a microscopic theory of the quasiparticles at the 
edge. In the low-temperature limit, 
the domination states contributing to the partition function 
are those states that the lowest Landau level of the CF-type excitations  
is fully filled in the bulk but only allow the edge CF-type excitations
to be gapless because the gap is shrinked in the edge due to the 
sharp edge potential. The other states with their energy $E_\gamma+E_g$ open 
a gap at least in the order of $\hbar \omega^*_c$ to the ground state. 
In the low-temperature limit, $k_BT\ll \hbar \omega^*_c$, the effective 
partition function is
\begin{eqnarray}
Z&\simeq&\sum_{\delta, N^e}C_N^{N^e} \int_{\partial} d^2z_1...d^2z_{N^e}
|\Psi_{e,\delta}|^2 e^{-\beta (E_\delta(N^e)+E_{g,b})}\\ \nonumber
&=&\sum_{N^e}C_N^{N^e}{\rm Tr_{(edge)}}e^{-\beta (H_e+E_{g,b})},
\label{apf}
\end{eqnarray}
where the trace runs over the low-lying set of the quantum state space for 
a fixed $N_e$ and, according to the single particle picture, 
$\Psi_{e,\delta}$ are the edge many-quasiparticle wave functions
. $E_\delta(N^e)$ is the eigen energy of the edge quasiparticle excitations
and $E_{g,b}$ is the
bulk state contribution to the ground state energy. For the disc 
sample, the edge quasiparticles are restricted in a circular strip near 
the boundary with its width $\delta R(\vec r)\ll R$ while the radius of the 
disc is $R$. The edge Hamiltonian of CFs reads
\begin{eqnarray}
H_e&=&\sum_{i=1}^{N^e}[\vec{p}_i-\vec{A}(\vec{r}_i)
+\vec{a}_e(\vec{r}_i)+\vec{a}_b(\vec{r}_i)]^2\\ \nonumber
&+&\sum_{i<j}V(\vec{r}_i-\vec{r}_j)+
\sum_i U_{eff}(\vec{r}_i),
\end{eqnarray}
where the external potential $U_{eff}$ is the effective potential 
including the interaction between the edge and bulk particles. We suppose the
potential is an infinity wall for $r\geq R$ and induces a constant 
electric field for $r<R$.
The statistics gauge field $\vec{a}$ is given by
\begin{eqnarray}
\vec{a}_e(\vec{r}_i)&=&\frac{\tilde\phi}{2\pi}\sum_{j\not=i}
\frac{\hat{z}\times (\vec{r}_{i}-\vec{r}_{j})}
{|\vec{r}_{i}-\vec{r}_{j}|^2},\\ \nonumber
\vec{a}_b(\vec{r}_i)&=&\frac{\tilde\phi}{2\pi}\sum_{a}
\frac{\hat{z}\times (\vec{r}_{i}-\vec{r}_{a})}
{|\vec{r}_{i}-\vec{r}_{a}|^2},
\end{eqnarray}
where $a$ is the index of the bulk electrons.
Taking the polar coordinate $x_i=r_i\cos\varphi_i,~ y_i=r_i\sin\varphi_i$,
the vector potential $A_\varphi(\vec{r}_i)=\frac{B}{2}r_i$ and $A_r(\vec{   
r_i})=0$. In the mean-field approximation, $a_{r,b}(\vec{r_i})=0$ and $a_ 
{\varphi}(\vec{r}_i)=B_{\tilde\phi}r_i/2$. Substituting the polar 
variations and the vector potential to
$H_e$ while using the mean-field value of $\vec{a}$, one has
\begin{eqnarray}
H_e&=&\sum_i\biggl[-\frac{\partial^2}{\partial r_i^2}+
(-\frac{i}{r_i}\frac{\partial}
{\partial \varphi_i}-\frac{B^*}{2}r_i)^2\\ 
\nonumber
&+&\frac{\tilde\phi^2}{4R^2}\sum_i(\sum_{j\not=i} 
\cot\frac{\varphi_{ij}}{2})^2\\ \nonumber
&-&
\frac{\tilde\phi}{R}\sum_{i<j}\cot\frac{\varphi_{ij}}{2}
\cdot i(\frac{\partial}{
\partial r_i}- \frac{\partial}{\partial r_j})-\frac{1}{R}
\frac{\partial}{\partial r_i}\biggr]\\ \nonumber
&+& V+U+O(\delta R/R).
\label{HE}
\end{eqnarray}

Now the eigen problem at the
edge could be solved by taking a trial wave function,
\begin{eqnarray}
\Psi_e(z_1,...,z_{N^e})&=&\exp\{\frac{it}{2}
\sum_{i<j}\frac{r_i-r_j}{R}\cot\frac{\varphi
_{ij}}{2}\}\\ \nonumber
&\times&f(r_1,...,r_{N^e})\Psi_s(\varphi_1,...,\varphi_{N^e}),
\label{TRI}
\end{eqnarray}  
where $t=\sqrt{\tilde\phi^2+\tilde\phi}-\tilde\phi$. The wave function $f$
is symmetric and $\Psi_s$ is anti-symmetric in the particle exchange. 
In the one-dimensional limit, $\delta R/R\to 0$ and a shift of the ground 
state energy, it is easy to 
see that the wave function $\Psi_s(\varphi_1,...,\varphi_{N^e})$ satisfies
the Schrodinger equation, $H_{s}\Psi_s=E_\varphi\Psi_s$
with
\begin{eqnarray}
H_s&=&H_{cs}+V, \label{HS}\\ 
H_{cs}&=&\sum_i(i\frac{\partial}{\partial x_i}+
\frac{B^*}{2}R)^2+\frac{g\pi^2}{L^2}\sum_{i<j}
\biggl[\sin(\frac{\pi x_{ij}}{L})\biggr]^{-2},
\label{HCS}
\end{eqnarray} 
where $x_{ij}=x_i-x_j$, $\varphi_i=\frac{2\pi x_i}{L}$ and  $L=2\pi R$ is 
the length of the boundary. The Hamiltonian (\ref{HCS}) is the CSLM 
Hamiltonian with a constant shift to the momentum operator. And the coupling 
constant 
\begin{equation}
g=2(\tilde\phi^2+\tilde\phi)=2(m^2-m), 
\end{equation}
where $m=\tilde\phi+1$ is an even integer.
So we see that the 
consistency between one- and two-dimensional  
statistics because the particle's
statistics in the CSLM \cite{Hald1} and anyons could be described by the same 
parameter $m$. The problem is exactly soluble and the wave functions are
known with the Bethe ansatz form or more exactly are given by the Jack 
polynomials \cite{EX}. The eigen-energy is
$E_\varphi=\sum_i (n_i-\frac{B^*}{2}R^2)^2/R^2$
where $n_i$ satisfy the Bethe ansatz equations
\begin{equation}
n_i=I_i+\frac{1}{2}\sum_{j\not=i}(m-1){\rm sgn}(n_i-n_j).
\label{BA}
\end{equation}

To see the low-lying excitations at the edge of the quantum Hall liquid 
droplet, we first turn off the interaction between electrons and 
will switch it on later.  
It is known that the CSLM has two-branches low-lying excitations, 
the left- and right-moving gapless sound waves. However, we will see
the magnetic field leads to the 
right-moving modes to be ranged out of the low-lying state sector. 
To see that, we substitute the azimuthal wave function to the trial wave 
function (\ref{TRI}) and then the radial wave function satisfies
\begin{equation}
\sum_i\biggl[-\frac{\partial^2 }{\partial r_i^2}
+(\frac{n_i}{r_i}-\frac{B^*}{2}r_i)^2\biggr]g
+U_{eff}g
+O(\frac{\delta R}{R})=Eg,
\end{equation}
with $g(r_1,..,r_{N^e})=e^{-r_i/2R}f$.
One finds that the many-body problem could be reduced to the single-particle
one except the $n_i$ are related by the Bethe ansatz equations (\ref{BA}).
Notice that it is consistent with the discussion to IQHE edge state 
if $\tilde\phi=0$ \cite{Halp}. We consider the external electric field lacks 
first. In the harmonic potential approximation,
one see that the single-particle wave function $g(r_n-r)$ with $r_n=\sqrt{
2|n|/B^*}$ satisfies
\begin{equation}
\biggl[-\frac{d^2}{dy^2}+\omega^{*2}_c y^2\biggr]g_+(y,s)
=\varepsilon g_+(y,s),
\label{abz}
\end{equation}
for $n\geq 0$, where $y=r-r_n$ and $s=R-r_n$ ( $O(\delta R/R)$ could give
$r_n$ a minor change). For $ n\leq 0$,
\begin{equation}
\biggl[-\frac{d^2}{dy^2}+\omega^{*2}_c y^2+|n|\omega^*_c\biggr]g_-(y,s)
=\varepsilon g_-(y,s).
\label{alz}
\end{equation}
Comparing (\ref{abz}) and (\ref{alz}), we see that there is a gap
$|n|\hbar\omega_c^*$ between states $g_+$ and $g_-$. The states with
the negative $n$ are ranged out of the low-lying state sector. The width of
the wave function $g_+$ is several times of $R_c^*$, the cyclotron radius of 
the CF in the effective field. (\ref{abz}) has been discussed by Halplerin 
years ago \cite{Halp} and one sees that the eigen-energy 
if $r_n=R$ is $\varepsilon^R_{n,\nu^*}=(2(\nu^*-1)+\frac{3}{2})
\hbar\omega^*_c$ since the wave function vanishes at $r=R$ 
while $\varepsilon_{n,\nu^*}=((\nu^*-1)+\frac{1}{2})\hbar\omega^*_c$
for $R-r_m\gg R_c^*$, a
harmonic oscillator energy and coinciding with the mean field theory applied
to the bulk states. The gapless excitations appear when $|R-r_n|\sim R_c^*$ .
Using a perturbative calculation, one has 
the bare exciation energy is $\varepsilon_0\sim v_c^*m^*\omega_c^*(r_n-R)$ 
with the cycltron velocity $v_c^*$ of a CF.
Now, turn the electric field on. A second order perturbative calculation
shows that \cite{Yu}
\begin{equation}
\varepsilon_0
=\delta\varepsilon_{n,\nu^*}=v_F^*(p_n-p_R)+\frac{1}{2m^*}(p_n-p_R)^2, 
\label{e0}
\end{equation}
where $p_n=m^*r_n\omega^*_c$ and the Fermi velocity $v_F^*=v_d^*+bv^*_c$
with $v_d^*=m\frac{cE}{B}$ and $b$ an order one constant. And $p_R=m^* R
\omega_c^*$, which shows the boundary of the droplet behaves like the Fermi 
surface \cite{Frad}. 
We see the linear
dispersion. Since $v_c^*\ll v^*_d$, the CF sound wave velocity is $v_F^*\approx
v_d^*=m v_d$ where $v_d$ is the CF drift velocity (the current velocity).
The relation between the current velocity and the sound wave velocity 
resembles the Haldane's velocity relations in the Luttinger liquid, 
$v_J=e^{2\varphi}v_s$, if we identify $m=e^{-2\varphi}$ \cite{Hald2}. Note 
that another zero point of (\ref{e0}) is at $p_n=p_R-2m^*v_F^*$ where
is actually deep inside of the bulk. 
So, we see that there is, indeed, only
one branch low-lying excitations ($\nu^*=1$) of the edge states, which behaves 
like the chiral Luttinger liquid (For details see \cite{Yu}). 

Now, let's make the relation to the macroscopic theory. In terms of the 
partition function (\ref{apf}), there is a most probable edge CF number
$\bar{N}^e$ which is given by $\delta Z/\delta N^e=0$.
$\bar{N}^e=\int dx \rho(x)$ with the edge density
$\rho(x)=h(x)\rho_e$ \cite{Wen}. Here $h(x)$ is the edge deformation and 
$\rho_e$ is the average density of the bulk electrons. 
One assumes the Fourier transformation of $h(x)=\sum_n e^{in\varphi}h_n$.
Then the Fourier components of the density is $\rho_n=h_n\rho_e$. Now, 
we identify $h_n=r_n-R$. One finds that 
\begin{equation}
\delta\varepsilon_{n,1}=2\pi mv_d\rho_n,
\end{equation}
and then the effective Hamiltonian reads
\begin{equation}
H_{eff}=\sum_{n>0} \rho_{-n}\delta\varepsilon_{n,1}=2\pi mv_d\sum_{n>0}
\rho_{-n}\rho_n,
\end{equation}
which is the Hamiltonian used by Wen from the hydrodynamic point of view
\cite{Wen}.

Now, let's switch on the interaction between CFs. First, we consider the 
short range interaction. In general, the CSLM pluses a short range interaction
is no longer soluble. So, we treat it as a perturbation to the CSLM. One sees 
that it varies the value of $n$ but does not renormalize the exponent
$m$ because the step-function phase shift is robust to the short range
perturbation \cite{Yu}. 
So the short range interaction 
does not affect the exponent $m$. This is consistent with the chiral
 Luttinger liquid
consideration \cite{Wen1}. The Coulomb interaction may take more space to 
discuss \cite{Yu}. Here we only give our recent result in a limit case. 
The one-dimensional Hamiltonian problem (\ref{HS}) for $N^e=2$ could 
be exactly
solved if $V=\frac{\alpha\pi}{L|\sin(\pi x_{ij}/L)|}$ in the large $L$ limit.
Actually, (\ref{HS}) is the same as the radial Hamiltonian of 
a three-dimensional 
electron scattering by a negative Coulomb interaction \cite{LL}. 
The phase shift, then, is given by
\begin{equation} 
\theta(k)={\rm sgn}(k)\tilde\phi\pi-2{\rm arg}\Gamma(m+i/k).
\end{equation}
By using the dilute gas limit, $x_1\ll x_2\ll...\ll x_{N^e}$, the many-body
problem could be solved asymptotically by the Bethe ansatz with the wave 
function
\begin{eqnarray}
&&\Phi_s(x_1,..,x_{N^e})=\sum_P A(P)\exp\{i\sum_ik_{P_i}x_i\\ \nonumber
&&-i\sum_{i<j}F(k_{P_i}-k_{P_j}, x_i-x_j)\}, \\ \nonumber
&&F(k,x)=\frac{\alpha}{2k}\ln(\biggl[\tan\frac{\pi x}{2L}\biggr]\frac{2\pi k}
{L}).
\end{eqnarray}
The asymptotic Bethe ansatz equation is given by
$
n_i=I_i+\frac{1}{2\pi}\sum_{j\not=i} \theta(k_i-k_j).
$
If we assume the Coulomb interaction is dramatically renormalized such 
that $\alpha L\ll 1$. In the thermodynamic limit and the zero temperature, 
$n_i$ could be solved by iteration and one has
$
n=\tilde n+C_n\ln(p_n-p_F),
$
where $\tilde n$ is regular if $p_n-p_F$ tends to zero and  $C_n$ is a constant
proportional to $\alpha L$. Hence, the radial equation reads
\begin{eqnarray}
&&-\frac{d^2 g_+(y,s)}{dy^2}+\omega^{*2}_c y^2 g_+(y,s)\\ \nonumber
&&+C'_n y\ln(p_n-p_R) g_+(y,s)
=\varepsilon g_+(y,s),
\end{eqnarray}
where $y=r-r_{\tilde n}$  and
$s=R-r_{\tilde n}$. We have identified $p_F=p_R$ here. By taking the 
approximation with $y\simeq s$ in the logarithm term, one see that the 
dispersion relation, after adding the external electric field,
\begin{equation}
\delta\varepsilon_{\tilde n,\nu^*}=(p_n-p_R)(v_d^*+A_n\ln(p-p_R)),
\end{equation}
with $A_n$ proportional to $\alpha L$.
We see that a branch of the excitations with the form $q\ln q$, which is
the plasma excitations caused by the Coulomb interaction \cite{Wen1}.   

Before concluding this paper, we would like to point out that the theory 
presented
here is only for the single branch edge excitations of the $\nu=1/m$ FQHE.
This procedure could be generalized to the many branch case, say $\nu=\frac{p}
{\tilde\phi p+1}$. The one-dimensional Hamiltonian for CFs is
\begin{equation}
H=\sum_{I,i_I}-\frac{\partial^2}{\partial x_{i_I}^2}+\frac{\pi^2}{2L^2}
\sum_{I,J;i_I\not=j_J}g_{IJ}\biggl[\sin \frac{\pi(x_{i_I}-x_{j_J})}{L}\biggr]
^{-2},
\end{equation} 
where $I,J=1,...,p$ are the branch indices and $g_{IJ}=2(K_{IJ}^2+K_{IJ})$. 
The matrix $K$ is given by $K_{IJ}=\tilde\phi+\delta_{IJ}$, which is 
introduced by Wen and Zee \cite{WZ}.

In conclusion, we have derived the microscopic model of the 
CF-type excitations for the edge excitations  
in the $\nu=1/m$  FQHE with the sharp edge by using the 
CF picture. The edge excitations are chiral because the magnetic
field suppresses one branch of the CSLM excitations.
The self-consistency between one-dimensional and two-dimensional statistics
is exhibited.

The authors are grateful to Z. B. Su and W. J. Zheng for helpful discussions. 
They would like to thank Y. S. Wu, for his important comments in the
earlier version of this paper and useful discussions. 
This work was supported in part 
by the NSF of China and the National PanDeng (Climb Up) Plan in China.

\end{document}